%% file: Main.tex
\documentclass[pdflatex,sn-mathphys]{sn-jnl}
\usepackage[T1]{fontenc}

\usepackage{graphicx}
\usepackage{fontawesome5}
\usepackage{xcolor}

\usepackage{amsmath}
\usepackage{amssymb}
\usepackage{enumitem}
\usepackage{tabularx}
\usepackage{multirow, makecell}
\usepackage{float}
\usepackage{breakcites}

\definecolor{idcolor}{HTML}{A6CE39}
\newcommand{\orcidlink}[1]{\href{https://orcid.org/#1}{\color{idcolor}\faOrcid}}

\jyear{2021}%

\theoremstyle{thmstyleone}%
\theoremstyle{thmstyletwo}%

\theoremstyle{thmstylethree}%

\begin{document}

\title[DAN using DBLP]{Deep Author Name Disambiguation using DBLP Data}


\author*[1,2]{\fnm{Zeyd} \sur{Boukhers}\orcidlink{0000-0001-9778-9164}}\email{zeyd.boukhers@fit.fraunhofer.de}

\author[2]{\fnm{Nagaraj Bahubali} \sur{Asundi}\orcidlink{0000-0002-1044-7047}}\email{nagarajbahubali@uni-koblenz.de}

\affil[1]{\orgdiv{Institute for Web Science and Technologies (WeST)}, \orgname{University of Koblenz-Landau}, \orgaddress{\street{Universitätsstraße 1}, \city{Koblenz}, \postcode{56070},  \country{Germany}}}

\affil[2]{\orgdiv{Department of Data Science and Artificial Intelligence}, \orgname{Fraunhofer Institute for Applied Information Technology -FIT-}, \orgaddress{\street{Schloss Birlinghoven 1}, \city{Sankt Augustin}, \postcode{53757},  \country{Germany}}}


\abstract{In the academic world, the number of scientists grows every year and so does the number of authors sharing the same names. Consequently, it challenging to assign newly published papers to their respective authors. Therefore, Author Name Ambiguity (ANA) is considered a critical open problem in digital libraries. This paper proposes an Author Name Disambiguation (AND) approach that links author names to their real-world entities by leveraging their co-authors and domain of research. To this end, we use data collected from the DBLP repository that contains more than 5 million bibliographic records authored by around 2.6 million co-authors. Our approach first groups authors who share the same last names and same first name initials. The author within each group is identified by capturing the relation with his/her co-authors and area of research, represented by the titles of the validated publications of the corresponding author. To this end, we train a neural network model that learns from the representations of the co-authors and titles. We validated the effectiveness of our approach by conducting extensive experiments on a large dataset.}


\maketitle

\input{Sources/introduction.tex}

\input{Sources/relatedwork.tex}

\input{Sources/method.tex}

\input{Sources/experiments.tex}

\input{Sources/conclusion.tex}

\bibliography{sn-bibliography}


\end{document}

%% file: Sources/introduction.tex
\section{Introduction}
\label{introduction}

AND is an important task in digital libraries that aims to properly link each publication to its respective co-authors so that author-level metrics can be accurately calculated and authors' publications can be easily found. However, this task is extremely challenging due to the high number of authors sharing the same names. In this paper, \emph{author name} denotes a sequence of characters referring to one or several authors~\footnote{It is estimated that about 114 million people share 300 common names.}, whereas \emph{author} refers to a unique person authoring at least one publication and cannot be identified only by his/her \emph{author name}~\footnote{In the DBLP database, there are 27 exact matches of ‘Chen Li’, 23 reverse matches and more than 1000 partial matches} but rather with the support of other identifiers such as ORCID, ResearchGate ID and Semantic Scholar author ID. 

Although relying on these identifiers almost eliminates any chance of mislinking a publication to its appropriate author, most bibliographic sources do not include such identifiers. This is because not all of the authors are keen to use these identifiers and if they are, there is no procedure or policy to include their identifiers when they are cited. Therefore, in bibliographic data (e.g. references), authors are commonly referred to by their names only. Considering the high number of authors sharing the same names (i.e. homonymy), it is difficult to link the names in bibliographic sources to their real-world authors especially when the source of the reference is not available or does not provide indicators of the author's identity. The problem is more critical when names are substituted by their initials to save space, and when they are erroneous due to wrong manual editing. Disciplines like social sciences and humanities suffer more from this problem as most of the publishers are small and mid-sized and cannot ensure the continuous integrity of the bibliographic data.  

Table~\ref{tab:illust} demonstrates real examples of reference strings covering the above-mentioned problems. The homonomy issue shows an example of two different papers citing the name \emph{J M Lee} which refers to two different authors. In this case, it is not possible to disambiguate the two authors without leveraging other features. The Synonymy issue shows an example of the same author \emph{Jang Myung Lee\orcidlink{0000-0003-4290-8087}} cited differently in two different papers as \emph{Jang Myung Lee} and \emph{J Lee}. Synonymy is a serious issue in author name disambiguation as it requires the awareness of all name variates of the given author. Moreover, some name variates might be shared by other authors, which increases homonymy. 


\begin{table}[ht]
    \caption{Illustrative examples of author name ambiguity and incorrect author names}
    \label{tab:illust}
    \centering
    \begin{tabular}{|c|c|m{8cm}|}
    
    \hline
        \textbf{Issue Type} &
        \textbf{Source} &
        \multicolumn{1}{c|}{\textbf{Citations}} \\
    \hline
    
    \multirow{3.7}{*}{Synonyms} 
        & See~\footnotemark 
        & T. Jin, \href{https://dblp.org/pid/130/8653.html}{\textbf{J. Lee}}, and H. Hashimoto, “Internet-based obstacle
        avoidance of mobile robot using a force-reflection,” in Proceedings of the 2004 IEEE/RSJ International Conference on
        Intelligent Robots and Systems, (Sendai, Japan), pp. 3418–
        3423, October 2004.\\
        
        \cline{2-3}
        
        & See~\footnotemark 
        & TasSeok Jin, \href{https://dblp.org/pid/130/8653.html}{\textbf{JangMyung Lee}}, and Hideki Hashimoto, “Internet-based
        obstacle avoidance of mobile robot using a force-reflection,” IEEE/RSJ International Conference on Intelligent Robots and Systems, pp. 3418-3423. 2004.\\
    \hline
    
    \multirow{4.5}{*}{Homonyms} 
        & See~\footnotemark 
        & T.S. Jin, \href{https://dblp.org/pid/130/8653.html}{\textbf{J.M. Lee}}, and H. Hashimoto. Internet-based obstacle
        avoidance of mobile robot using a force-reflection. In Proceedings
        of the 2004 IEEE/RSJ International Conference on Intelligent Robots
        and Systems, pages 3418–3423, Sendai, Japan, October 2004.\\
        
        \cline{2-3}
        
        & See~\footnotemark  
        & H-J Kim, \href{https://dblp.org/pid/53/6517.html}{\textbf{J-M Lee}}, J-A Lee, S-G Oh, W-Y Kim, "Contrast Enhancement
        Using Adaptively Modified Histogram Equalization", Lecture Notes in
        Computer Science, Vol.4319, pp.1150 - 1158, Dec. 2006.\\ 
    \hline

    \end{tabular}
\end{table}

\addtocounter{footnote}{-3}
\footnotetext{Xu, Zhihao, et al. "Teleoperating a formation of car-like rovers under time delays." Proceedings of the 30th Chinese Control Conference. IEEE, 2011.}
\addtocounter{footnote}{1}
\footnotetext{Shi, Pu, Jianning Hua, and Yiwen Zhao. "Posture-based virtual force feedback control for teleoperated manipulator system." 2010 8th World Congress on Intelligent Control and Automation. IEEE, 2010.}
\addtocounter{footnote}{1}
\footnotetext{Xu, Zhihao, Lei Ma, and Klaus Schilling. "Passive bilateral teleoperation of a car-like mobile robot." 2009 17th Mediterranean Conference on Control and Automation. IEEE, 2009.}
\addtocounter{footnote}{1}
\footnotetext{
Lu, Ching-Hsi, Hong-Yang Hsu, and Lei Wang. "A new contrast enhancement technique by adaptively increasing the value of histogram." 2009 IEEE international workshop on imaging systems and techniques. IEEE, 2009.}

Since these problems are known for decades, several studies~\cite{muller2017semantic,kim2018web,foxcroft2019name2vec,hussain2017survey,ferreira2012brief,qian2015dynamic,zhang2016bayesian,khabsa2014large,khabsa2015online} have been conducted using different machine learning approaches. This problem is often tackled using supervised approaches such as Support Vector Machine (SVM)~\cite{han2004two}, Bayesian Classification~\cite{zhang2016bayesian} and Neural networks (NN)~\cite{tran2014author}. These approaches rely on the matching between publications and authors which are verified either manually or automatically. Unsupervised approaches~\cite{liu2014author,kim2020learning,fan2011graph} have also been used to assess the similarity between a pair of papers. Other unsupervised approaches are also used to estimate the number of co-authors sharing the same name~\cite{zhang2018name} and decide whether new records can be assigned to an existing author or a new one~\cite{qian2015dynamic}. Due to the continuous increase of publications, each of which cites tens of other publications and the difficulty to label this streaming data, semi-supervised approaches~\cite{louppe2016ethnicity,zhao2013semi} were also employed. Recent approaches~\cite{zhang2017name,xu2018network} leveraged the outstanding efficiency of deep learning on different domains to exploit the relationship among publications using network embedding. All these approaches use the available publication data about authors such as titles, venues, year of publication and affiliation. Some of these approaches are currently integrated into different bibliographic systems. However, all of them require an exhausting manual correction to reach an acceptable accuracy. In addition, most of these approaches rely on the metadata extracted from the papers which are supposed to be correct and complete. In real scenarios, the source of the paper is not always easy to find and only the reference is available.

In this paper, which builds upon our earlier work~\cite{boukhers2022whois}, we aim to employ bibliographic data consisting of publication records to link each author's name in unseen records to their appropriate real-world authors (i.e. DBLP identifiers) by leveraging their co-authors and area of research embedded in the publication title and source. Note that the goal of this paper is to disambiguate author names in newly published papers that are not recorded in any bibliographic database. Therefore, all records that are considered unseen are discarded from the bibliographic data and used only for testing the approach. The assumption is that any author is most likely to publish articles in specific fields of research. Therefore, we employ articles' titles and sources (i.e. Journal, Booktitle, etc.) to bring authors close to their fields of research represented by the titles and sources of publications. We also assume that authors who already published together are more likely to continue collaborating and publishing other papers.

For the goal mentioned above, our proposed model \emph{WhoIs} is trained on a bibliographic collection obtained from DBLP, where a sample consists of a target author, pair of co-authors, title and source. For co-authors, the input is a vector representation obtained by applying Char2Vec which returns character-level embedding of words. For title and source, the BERT model is used to capture the semantic representations of the sequence of words. Our model is trained and tested on a challenging dataset, where thousands of authors share the same atomic name variate. The main contributions of this paper are: 
\begin{itemize}[leftmargin=*]

\item We proposed a novel approach for author name disambiguation using semantic and symbolic representations of titles, sources, and co-authors.
\item We provided a statistical overview of the problem of author name ambiguity. 
\item We conducted experiments on challenging datasets simulating a critical scenario. 
\item The obtained results and the comparison against baseline approaches demonstrate the effectiveness of our model in disambiguating author names.
\end{itemize}

The rest of the paper is organized as follows. Section~\ref{related_work} briefly presents related work. Section~\ref{method} describes the proposed framework. Section~\ref{experiments} presents the dataset, implementation details and the obtained results of the proposed model. Finally, Section~\ref{conclusion} concludes the paper and gives insights into future work.

%% file: Sources/relatedwork.tex
\section{Related Work}
\label{related_work}

In this section, we discuss recent approaches softly categorized into three categories, namely unsupervised-, supervised- and graph-based;

\subsection{Unsupervised-based:} Most of the studies treat the problem of author name ambiguity as an unsupervised task~\cite{kim2020learning,zhang2018name,khabsa2015online,khabsa2015online,qian2015dynamic} using algorithms like DBSCAN~\cite{khabsa2015online} and agglomerative
clustering~\cite{wu2014unsupervised}. Liu et al.~\cite{liu2014author} and Kim et al.~\cite{kim2020learning} rely on the similarity between a pair of records with the same name to disambiguate author names on the PubMed dataset. Zhang et al.~\cite{zhang2018name} used Recurrent Neural Network (RNN) to estimate the number of unique authors in the Aminer dataset. This process is followed by manual annotation. In this direction, Ferreira et al.~\cite{ferreira2010effective} have proposed a two-phase approach applied to the DBLP dataset, where the first one is obtaining clusters of authorship records and then disambiguation is applied to each cluster. Wu et al.~\cite{wu2014unsupervised} fused features such as affiliation and content of papers using Shannon’s entropy to obtain a matrix representing pairwise correlations of papers which is in return used by Hierarchical Agglomerative Clustering (HAC) to disambiguate author names on Arnetminer dataset. Similar features have been employed by other approaches~\cite {yang2011author,arif2014author}.

\subsection{Supervised-based:} 
Supervised approaches~\cite{han2004two,qian2011combining,sun2011detecting,tran2014author,zhang2016bayesian} are also widely used but mainly only after applying to block that gathers authors sharing the same names together. Han et al.~\cite{han2004two} present two supervised learning approaches to disambiguate authors in cited references. Given a reference, the first approach uses the Naive Bayes model to find the author class with the maximal posterior probability of being the author of the cited reference. The second approach uses SVM to classify references from DBLP to their appropriate authors. Sun et al.~\cite{sun2011detecting} employ heuristic features like the percentage of citations gathered by the top name variations for an author to disambiguate common author names. Neural networks are also used~\cite{tran2014author} to verify if two references are close enough to be authored by the same target author or not. Hourrane et al.~\cite{hourrane2018using} propose a corpus-based approach that uses word embeddings to compute the similarity between cited references. In~\cite{ebraheem2018distributed}, an Entity Resolution system called the DEEPER is proposed. It uses a combination of bi-directional recurrent neural networks (BRNN) along with Long Short Term Memory (LSTM) as the hidden units to generate a distributed representation for each tuple to capture the similarities between them. Zhang et al.~\cite{zhang2016bayesian} proposed an online Bayesian approach to identify authors with ambiguous names and as a case study, bibliographic data in a temporal stream format is used and the disambiguation is resolved by partitioning the papers into homogeneous groups.

\subsection{Graph-based:}
As bibliographic data can be viewed as a graph of citations, several approaches have leveraged this property to overcome the problem of author name ambiguation~\cite{hoffart2011robust,han2011collective,zhang2017name,xu2018network}. Hoffart et al.~\cite{hoffart2011robust} present a method for collective disambiguation of author names, which harnesses the context from a knowledge base and uses a new form of coherence graph. Their method generates a weighted graph of the candidate entities and mentions to compute a dense sub-graph that approximates the best entity-mention mapping.  Xianpei et al.~\cite{han2011collective} aim to improve the traditional entity linking method by proposing a graph-based collective entity linking approach that can model and exploit the global interdependence, i.e., the mutual dependence between the entities. In~\cite{zhang2017name}, the problem of author name ambiguity is overcome using relational information considering three graphs: person-person, person-document and document-document. The task becomes then a graph clustering task with the goal that each cluster contains documents authored by a unique real-world author. For each ambiguous name, Xu et al.~\cite{xu2018network} build a network of papers with multiple relationships. A network-embedding method is proposed to learn paper representations, where the gap between positive and negative edges is optimized. Further, HDBSCAN is used to cluster paper representations into disjoint sets such that each set contains all papers of a unique real-world author.

%% file: Sources/method.tex
\section{Approach: }
\label{method}

In this paper, AND is designed using a bibliographic dataset $\mathcal{D}=\{d_i\}_{i=1}^{N}$, consisting of $N$ bibliographic records, where each record $d_i$ refers to a unique publication such that $d_i=\{t_i, s_i, \langle a_{i,u},\delta_{i,u}\rangle _{u=1}^{\omega_i}\}$. Here, $t_i$ and $s_i$ denote the \emph{title} and \emph{source} of the record, respectively. $a_{i,u}$ and $\delta_{i,u}$ refer to the $u$\emph{th} author and its corresponding name, respectively, among $\omega_i$ co-authors of $d_i$. Let $\Delta=\{\delta(m)\}_{m=1}^{M}$ be a set of $M$ unique author names in $D$ shared by a set of $L$ unique authors $\mathcal{A}=\{a(l)\}_{l=1}^{L}$ co-authoring all records in $D$, where $L>>M$. Note that each author name $\delta(m)$ might refer to one or more authors in $\mathcal{A}$ and each author $a(l)$ might be referred to by one or two author names in $\Delta$. This is because we consider two variates for each author as it might occur differently in different papers. For example, the author ``\emph{Rachid Deriche}''\orcidlink{0000-0002-4643-8417} is assigned to two elements in $\Delta$, namely ``\emph{Rachid Deriche}'' and ``\emph{R. Deriche}''.

\begin{figure*}[h]
  \centering
  \includegraphics[width=0.85 \linewidth]{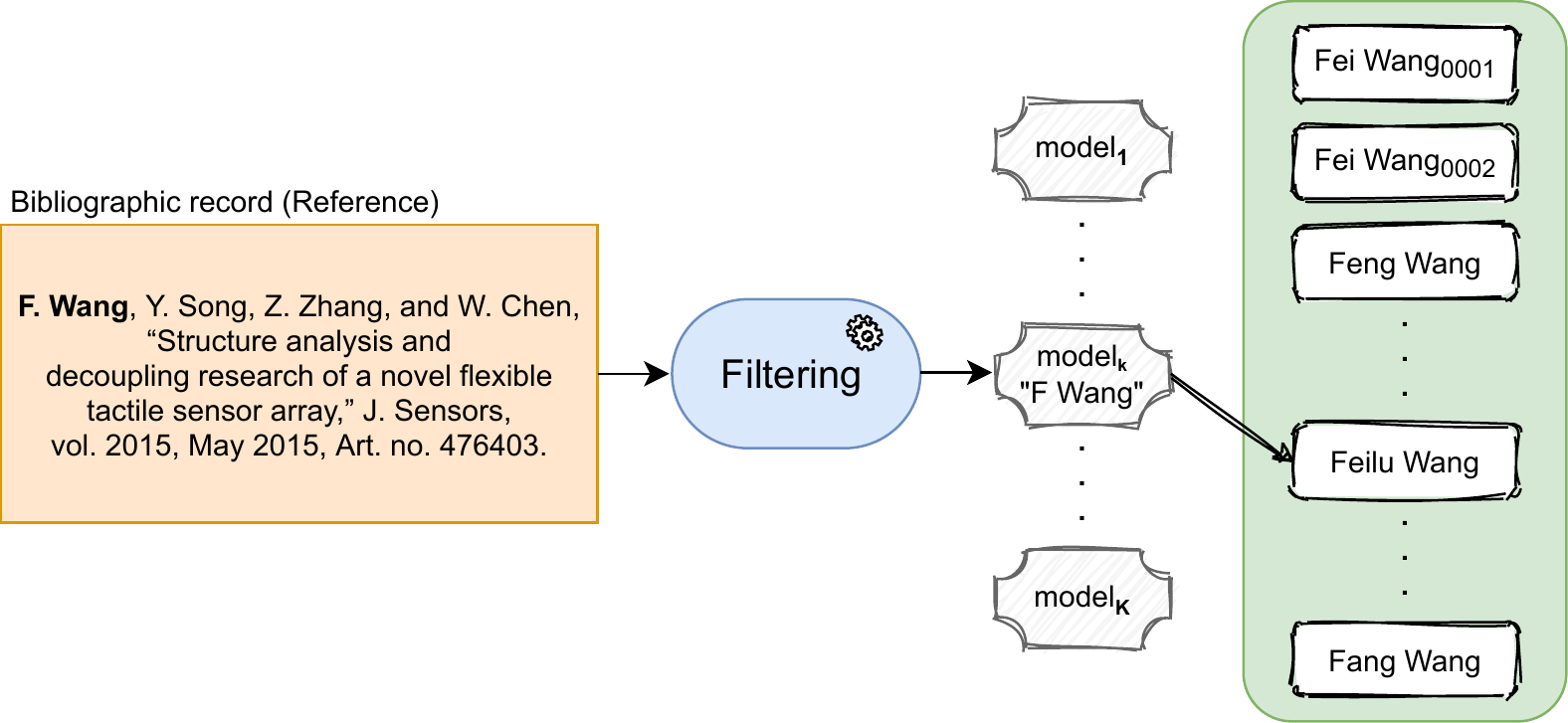}
  \caption{An illustration of the task for linking a name mentioned in the reference string with the corresponding DBLP author entity.}
  \label{fig:illust}
\end{figure*}

Given a reference record $d^* \notin \mathcal{D}$, the goal of our approach is to link each author name $\delta^*_{u} \in \Delta$ that occurs in $d^*$ to the appropriate author in $\mathcal{A}$ by leveraging $t^*$, $s^*$ and $\{\delta^*_{u}\}_{u=1}^{\omega^*}$. Figure~\ref{fig:illust} illustrates an overview of our proposed approach. First, the approach computes the correspondence frequency $\delta^*_{u}\mathbf{R}\mathcal{A}$ that returns the number of authors in $\mathcal{A}$ corresponding to $\delta^*_{u}$. $\delta^*_{u}\mathbf{R}\mathcal{A} = 0$ indicates that $\delta^*_{u}$ corresponds to a new author $a(\textrm{new}) \notin \mathcal{A}$. $\delta^*_{u}\mathbf{R}\mathcal{A} = 1$ indicates that $\delta^*_{u}$ corresponds to only one author $a(l) \in \mathcal{A}$. In this case, we directly assign $\delta^*_{u}$ to $a(l)$ and no further processing is necessary. Note that in this case, $\delta^*_{u}$ might also refer to a new author $a(\textrm{new}) \notin \mathcal{A}$ who has the same name as an existing author $a(l) \in \mathcal{A}$. However, our approach does not handle this situation. Please refer to Section~\ref{subsec:limit} that lists the limitation of the proposed approach.

The goal of this paper is to handle the case of  $\delta^*_{u}\mathbf{R}\mathcal{A} > 1$ which indicates that $\delta^*_{u}$ can refer to more than one author. To this end, the approach extracts the atomic name variate from the author name $\delta^*_{u}$. For example, for the author name $\delta^*_{u} = $ ``\emph{Lei Wang}'', the atomic name variate is $\overline{\delta^*_{u}} = $  ``\emph{L Wang}''. Let $\overline{\delta^*_{u}}$ correspond to $\overline{\delta_\mu}$ which denotes the $\mu$\emph{th} atomic name variate among $K$ possible name variates. Afterwards, the corresponding Neural Network model $\theta_\mu \in \Theta=\{\theta_k\}_{k=1}^{K}$ is picked to distinguish between all authors $\mathcal{A}_\mu=\{a(l_\mu)\}_{l_\mu=1}^{L_\mu}$ who share the same name variate $\overline{\delta_\mu}$.\\

  \vspace{-0.4cm}

\subsection{Model Architecture}

\begin{figure*}[t!]
  \centering
  \includegraphics[width=0.85 \linewidth]{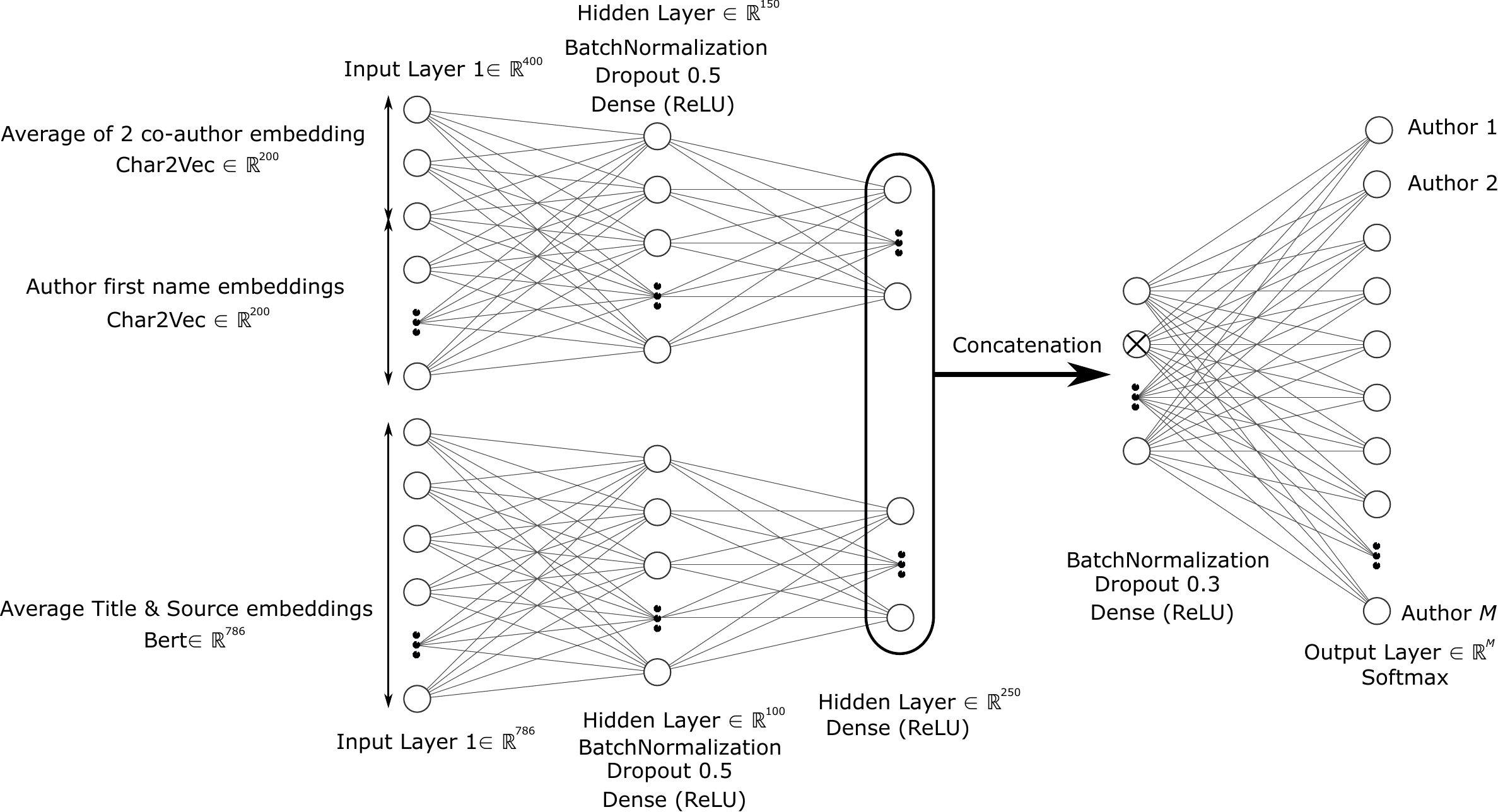}
  \caption{The architecture of our model.}
  \label{fig:arch}
  \vspace{-0.5cm}
\end{figure*}

The Neural Network (NN) model $\theta_\mu$ takes as input the attributes of $d^*$, namely the first name of the target author $\delta^{* \textrm{first-name}}_{u}$, full names of two co-authors $\delta^{*}_{p}$ and $\delta^{*}_{j}$, title $t^*$ and source $s^*$. Figure~\ref{fig:arch} illustrates the architecture of $\theta_\mu$, with an output layer of length $L_k$ corresponding to the number of unique authors in $\mathcal{A_\mu}$ who have the same atomic name variate $\delta_k$. As shown in Figure~\ref{fig:arch}, $\theta_\mu$ takes two inputs $\mathbf{x_{\mu,1}}$ and $\mathbf{x_{\mu,2}}$, such that:

\begin{equation}
\centering
 \begin{split}
    \mathbf{x_{\mu,1}}= \textrm{char2vec}(\delta^{* \textrm{first-name}}_{u})\bigoplus \frac{1}{2}\left( \textrm{char2vec}(\delta^{*}_{p}) + \textrm{char2vec}(\delta^{*}_{j})\right), \\
    \mathbf{x_{\mu,2}}=\frac{1}{2}\left( \textrm{bert}(t^*) + \textrm{bert}(s^*)\right),
    \end{split}
\end{equation}

\noindent where $\textrm{char2vec}(\mathbf{w})$ returns a vector representation of length $200$ generated using \emph{Char2Vec}~\cite{cao2016joint}, which provides a symbolic representation of $w$.  $\textrm{bert}(\mathbf{w})$ returns a vector representation of each token in $\mathbf{w}$ w.r.t its context in the sentence. This representation of length $786$ is generated using BERT~\cite{devlin2018bert}. The goal of separating the two inputs is to overcome the sparseness of content embedding and force the model to emphasise more on target author representation.

All the hidden layers possess a ReLU activation function, whereas the output is a Softmax classifier. Since the model has to classify thousands of classes, each of which is represented with very few samples, $50\%$ of the units in the last hidden layers are dropped out during training to avoid over-fitting. Furthermore, the number of publications significantly differs from one author to another. Therefore, each class (i.e. the author) is weighted according to its number of samples (i.e. publications). The model is trained with \emph{adam} optimizer and sparse categorical cross-entropy loss function. Our empirical analysis showed that the best performance was achieved with this architecture and these parameters, which were obtained through grid search.

\subsection{Author name representation}

The names of authors do not hold any specific semantic nature as they are simply a specific sequence of characters referring to one or more persons. Therefore, we need a model that can encode words based on the order and distribution of characters such that author names with a similar name 
spellings are encoded closely, assuming possible manual editing errors of cited papers. 

Chars2vec is a powerful NN-based language model that is preferred when the text consists of abbreviations, typos, etc. It captures the non - vocabulary words and places words with similar spelling closer in the vector space. This model uses a fixed list of characters for word vectorization, where a one-hot encoding represents each character.

\subsection{Source and Title embedding}

The source (e.g. journal names and book titles) of reference can provide a hint about the area of research of the given reference. In addition, the title is a meaningful sentence that embeds the specific topic of the reference. Therefore, we used these two features to capture the research area of the author. Contrary to the author's name, the goal here is to capture the context of the sequences of words forming the title and source. Therefore, we employed the pre-trained BERT model~\cite{devlin2018bert} to obtain sentence embeddings of both the title and source.

\subsection{Model Training}

Given the training set $\mathcal{D}_\mu \subset \mathcal{D}$ that corresponds to the subset of bibliographic records authored by authors having the atomic name variate $\overline{\delta_\mu}$, $d_{i_\mu} \in \mathcal{D}_\mu$ generates $\omega_{i_\mu}$ training samples $\langle \delta_\mu, \delta_{i_\mu,p}, \delta_{i_\mu,j}, t_{i_\mu}, s_{i_\mu} \rangle_{p=1}^{\omega_{i_\mu}}$, where $\delta_{i_\mu,j}$ is a random co-author of $d_{i_\mu}$ and might be also the same author name as $\delta_{i_\mu,p}$ and/or $\delta_\mu$. Note also that we consider one combination where $\delta_{i_\mu,p} = \delta_\mu$. In order to train the model with the other common name variate where the first name is substituted with its initial, for each sample, we generate another version with name variates $\langle \overline{\delta_\mu}, \overline{\delta_{i_\mu,p}}, \overline{\delta_{i_\mu,j}}, t_{i_\mu}, s_{i_\mu} \rangle$. Consequently, each bibliographic record is fed into the model $2 \times \omega_{i_\mu}$ times. 

Since the third co-author $\delta_{i_\mu,p}$ is randomly assigned to the training sample among $\omega_{i_\mu}$ co-authors  $d_{i_\mu}$, we randomly reassign it after $Y$ epochs. In addition to lower training complexity, this has shown in the conducted experiments a slightly better result than training the model at each epoch with samples of all possible co-author pairs $p$ and $j$. 

\subsection{Model Tuning}

For each training epoch, \emph{WhoIs} model fine-tunes the parameters to predict the appropriate target author. The performance of the model is considerably influenced by the number of epochs set to train. Specifically, a low epoch count may lead to underfitting. Whereas, a high epoch count may lead to over-fitting. To avoid this, we enabled early stopping, which allows the model to specify an arbitrarily large number of epochs.

Keras supports early stopping of the training via a callback called \emph{EarlyStopping}. This callback is configured with the help of the \emph{monitor} argument which allows setting the validation loss. With this setup, the model receives a trigger to halt the training when it observes no more improvement in the validation loss.

Often, the very first indication of no more improvement in the validation loss would not be the right epoch to stop training; because the model may start improving again after passing through a few more epochs. We overcome this by adding a delay to the trigger in terms of consecutive epochs count on which, we can wait to observe no more improvement. A delay is added by setting the \emph{patience} argument to an appropriate value. \emph{patience} in \emph{WhoIs} is set to $50$, so that the model only halts when the validation loss stops getting better for the past 50 consecutive epochs.

\subsection{Model checkpoint}
Although \emph{WhoIs} stops the training process when it achieves a minimum validation loss, the model obtained at the end of the training may not give the best accuracy on validation data. To account for this, Keras provides an additional callback called \emph{ModelCheckpoint}. This callback is configured with the help of another \emph{monitor} argument. We have set the \emph{monitor} to monitor the validation accuracy. With this setup, the model updates the weights only when it observes better validation accuracy compared to earlier epochs. Eventually, we end up persisting in the best state of the model with respect to the best validation accuracy.

\subsection{Prediction:}
Given the new bibliographic record $d^*=\{t^*, s^*, \langle \delta^*_{u}\rangle _{u=1}^{\omega^*}\}$, the goal is to disambiguate the author name $\delta^*_{\textrm{target}}$ which is shared by more than one author ($\delta^*_{\textrm{target}}\mathbf{R}\mathcal{A} > 1$). To this end, $Y$ samples $S_{y=1}^Y$ are generated for all possible pairs of co-author names $p$ and $j$: $\langle \delta^*_{\textrm{target}}, \delta^*_{p}, \delta^*_{j}, t^*, s^* \rangle_{p=1,j=1}^{\omega^*,\omega^*}$, where  $Y = \textrm{C}(\omega^*+1, 2)$, i.e. the combination of $\omega^*+1$ authors taken 2 at a time, and $\delta^*_u$ can be a full or abbreviated author name. All the $Y$ samples are fed to the corresponding model $\theta_\mu$, where the target author $a_{\textrm{target}}$ of the target name $\delta^*_{\textrm{target}}$ is predicted as follows:

\begin{equation}
    a_{{target}}=\underset{1\cdots L_\mu}{{argmax}} \left(\theta_\mu(S_1)\oplus \theta_\mu(S_2)\oplus\cdots\oplus \theta_\mu(S_Y)\right),
\end{equation}

\noindent where $\theta_\mu(S_y)$ returns a probability vector of length $L_\mu$ with each element $l_\mu$ denotes the probability of the author name $\delta^*_{\textrm{target}}$ to be the author $a_{l_\mu}$.

%% file: Sources/experiments.tex
\section{Experiments}
\label{experiments}

This section presents the experimental results of the proposed approach to the DBLP dataset.

\subsection{Dataset}

The following datasets are widely used to evaluate author name disambiguation approaches but the results on these datasets cannot reflect the results on real scenario streaming data.

\begin{itemize}[leftmargin=*]
    \item \textbf{ORCID~\footnote{\url{https://figshare.com/articles/ORCID_Public_Data_File_2017/5479792}}:} it is the largest accurate dataset as the publication is assigned to the author only after authorship claim or another rigorous authorship confirmation. However, this accuracy comes at the cost of the number of assignments. Our investigation shows that most of the registered authors are not assigned to any publication and an important number of authors are not even registered. This is because most of the authors are not keen to claim their publications due to several reasons.  
    
    \item \textbf{KDD Cup 2013~\footnote{\url{https://www.kaggle.com/c/kdd-cup-2013-author-paper-identification-challenge}}:} it is a large dataset that consists of 2.5M papers authored by 250K authors. All author metadata are available including affiliation. 
    
    \item \textbf{Manually labelled (e.g. PENN~\footnote{\url{http://clgiles.ist.psu.edu/data/nameset_author-disamb.tar.zip}}, QIAN~\footnote{\url{https://github.com/yaya213/DBLP-Name-Disambiguation-Dataset}}, AMINER~\footnote{\url{http://arnetminer.org/lab-datasets/disambiguation/rich-author-disambiguation-data.zip}}, KISTI~\footnote{\url{http://www.lbd.dcc.ufmg.br/lbd/collections/disambiguation/DBLP.tar.gz/at_download/file}}):} These datasets are supposed to be very accurate since they are manually labelled. However, this process is expensive and time-consuming and, therefore, it can cover only a small portion of authors who share the same names. 
    
\end{itemize}

In this work, we collected our dataset from the DBLP bibliographic repository\footnote{\url{https://dblp.uni-trier.de/xml/} (July 2020)}. The DBLP version of July 2020 contains 5.4 million bibliographic records such as conference papers, articles, thesis, etc., from various fields of research. 
 As stated by the maintainers of DBLP~\footnote{\url{https://dblp.org/faq/How+accurate+is+the+data+in+dblp.html}}, the accuracy of the data is not guaranteed. However, a lot of effort is put into manually disambiguating homonym cases when reported by other users. Consequently, we are aware of possible homonym cases that are not resolved yet. 
 From the repository, we collected only records of publications published in journals and proceedings. Each record in this collection represents metadata information of a publication with one or more authors, title, journal, year of publication and a few other attributes. The availability of these attributes differs from one reference to another. Also, the authors in DBLP who share the same name have a suffix number to differentiate them. For instance, the authors with the same name ‘Bing Li’ are given suffixes such as ‘Bing Li 0001’, and ‘Bing Li 0002’. The statistical details of the used DBLP collection are shown in Table~\ref{tab:dblp_stat}.

\begin{table}[]
    \caption{Statistical details of the used DBLP collection. }
    \label{tab:dblp_stat}
    \centering
    \begin{tabular}{|l|c|}
    \hline
    \# of records     &  $5258623$\\
    \hline
    \# of unique authors & $2665634$\\
    \hline
    \# of unique author names & $2613577$\\
     \hline
    \# of unique atomic name variates & $1555517$\\
    \hline
    \end{tabular}
    \vspace{-0.3cm}
\end{table}

\begin{figure}[ht]
\centering
  \includegraphics[width=0.8\textwidth]{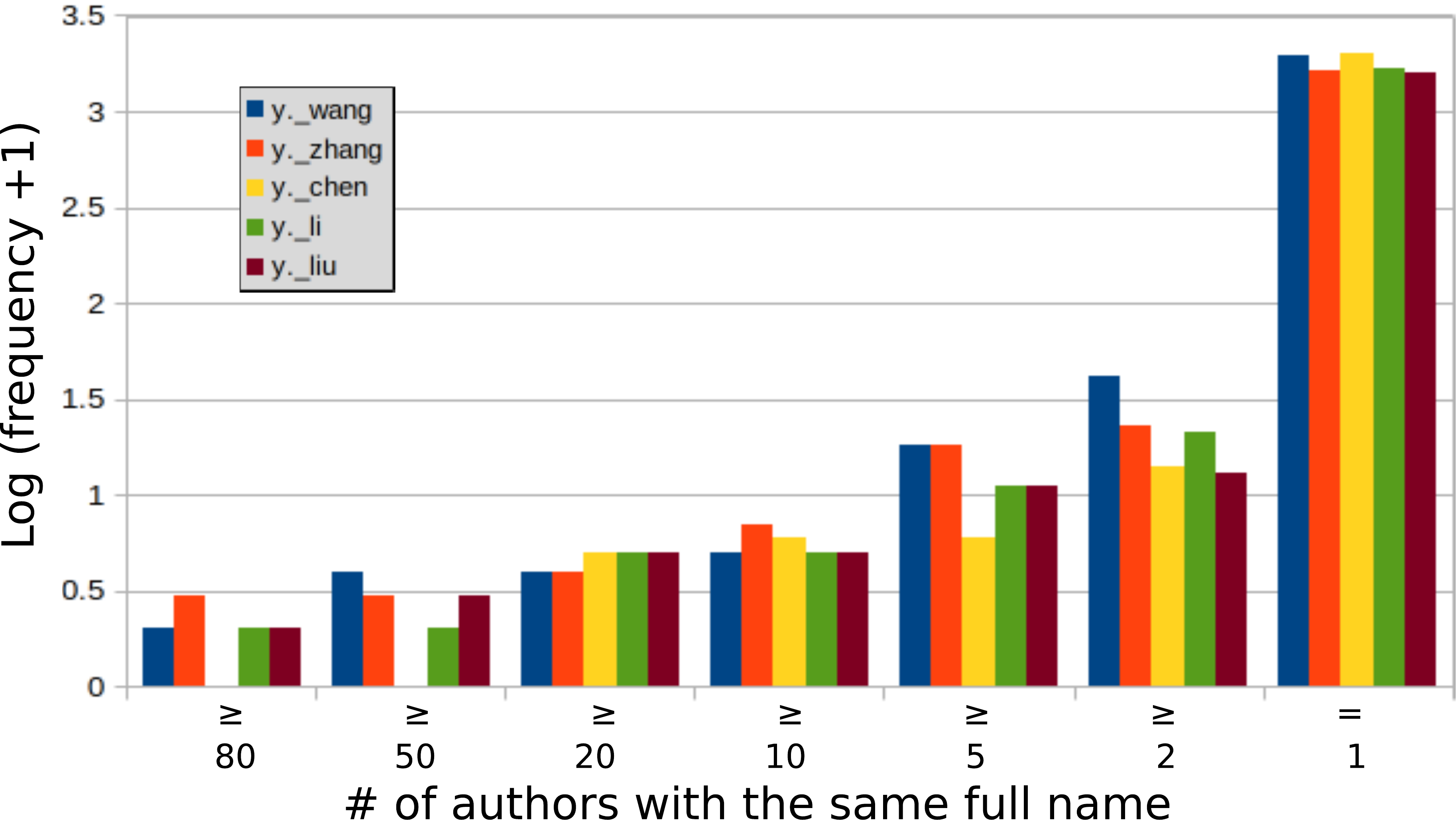}
  \caption{The \emph{log} frequency of authors sharing the same full name for the top five sub-collections.}
  \label{fig:authrec}
\end{figure}

Figure~\ref{fig:authrec} indicates that the majority of target authors in the sub-collections (each sub-collection includes all records of authors with the same name) have distinct full names. However, a considerable number of them share full names, leading to a significant challenge, particularly when multiple authors (e.g. over 80 in 4 out of 5 sub-collections) share the same full name but have an unequal number of publications. In such cases, it becomes more challenging to differentiate these authors from the dominant author with the same name.

\begin{figure}[ht]
\centering
  \includegraphics[width=0.8\textwidth]{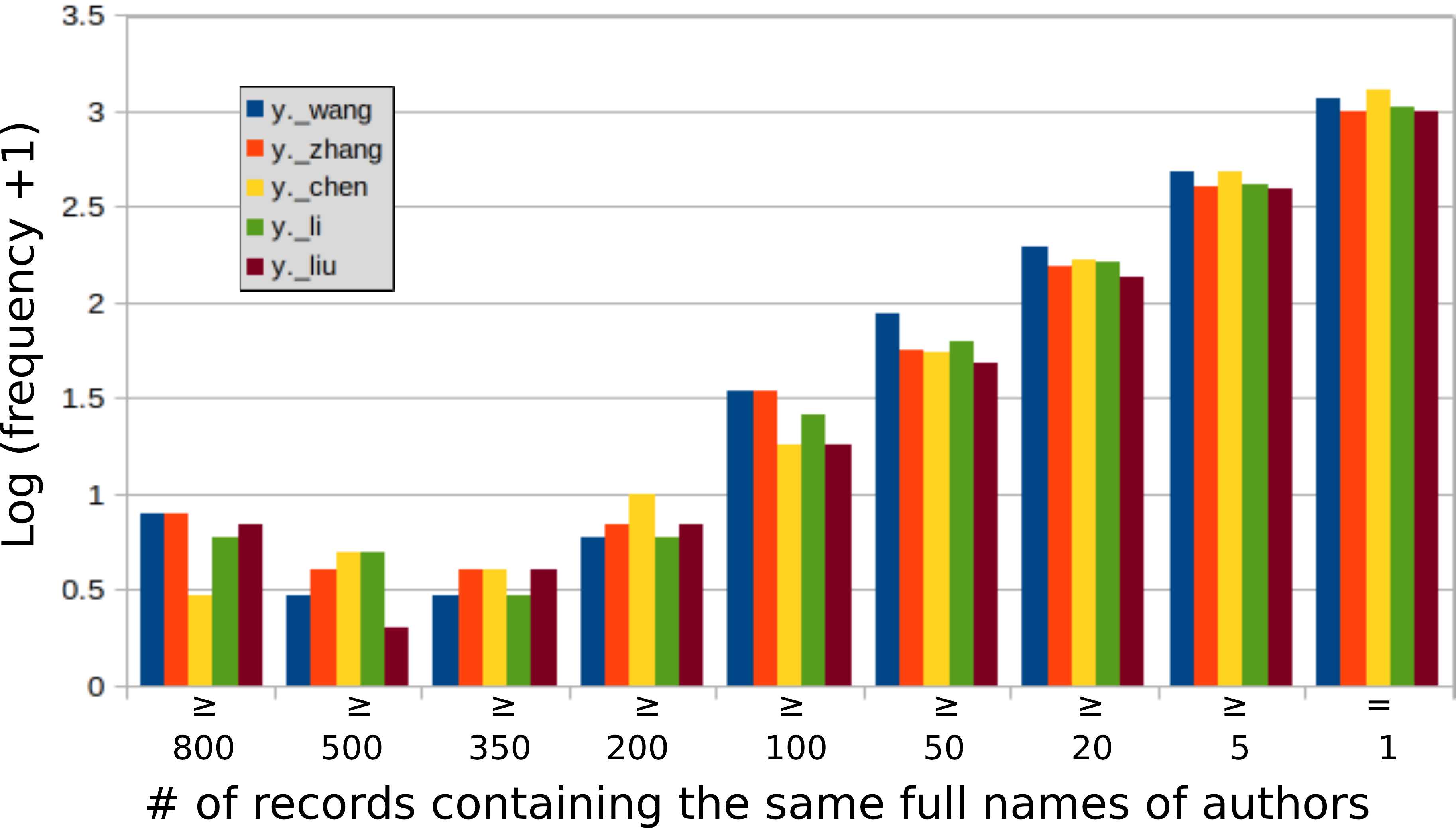}
  \caption{The \emph{log} frequency of records with the same full name of the target author for the top five sub-collections.}
  \label{fig:authname}
\end{figure}

Figure~\ref{fig:authname} illustrates the log frequency of bibliographic records with the same full name in the top five sub-collections used in this paper. As illustrated, in all sub-collections, the target authors of around half of the records authored a few records (less than 5) and have unique names. Although it is simple to distinguish these authors when their full names occur, it is extremely challenging to recognize them among more than $2000$ authors sharing the same atomic name variate due to the unbalance of records with the other authors.

\begin{figure}[ht]
    \centering
    \includegraphics[scale=0.2]{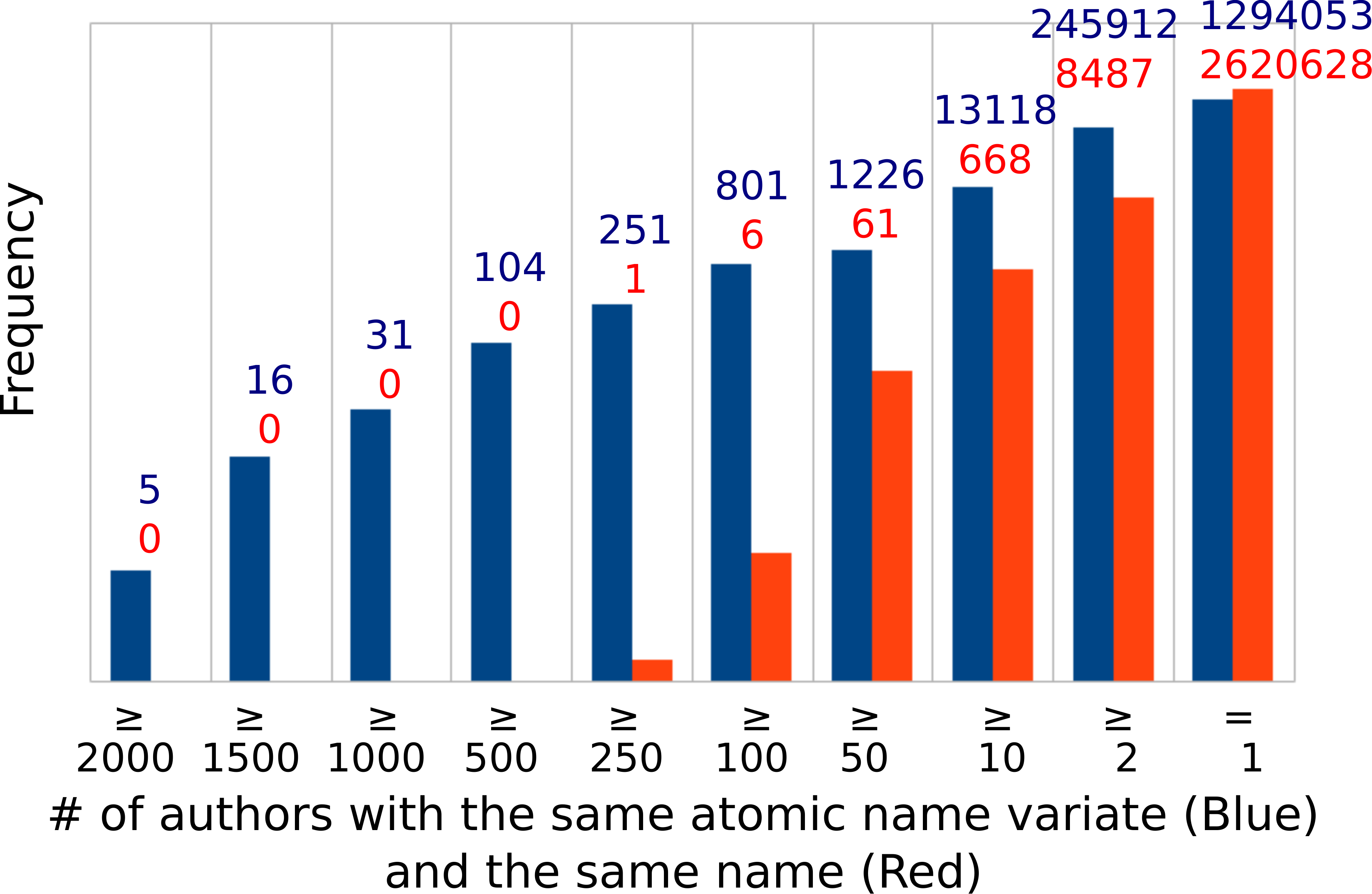}
    \caption{Frequency of authors sharing the same atomic name variate (Blue) / the same full name (Red).}
    \label{exp:frequency}
\end{figure}

Figure~\ref{exp:frequency} shows the frequency of authors sharing the same names and the same atomic name variates. As can be seen, the problem is more critical when the authors are cited with their atomic name variate as there are five atomic name variates shared by around $11.5k$ authors. This makes the problem of disambiguation critical because not only targets authors who might share the same atomic name variate but also their co-authors. For instance, we observed publications authored by the pair of co-authors having the atomic name variates: \emph{Y. Wang} and \emph{Y. Zhang}. However, they refer to different \emph{Y. Wang} and \emph{Y. Zhang} pairs of real-world authors.

\vspace{-0.5cm}

Since our approach gathers authors with the same name variates, $261464$ models are required to disambiguate all author names in our collection. Therefore, we present in this paper the experimental results on 5 models corresponding to the highest number of authors sharing the same name variates. Table~\ref{tab:stat} presents statistical details of the five sub-collections which demonstrates the challenges inherent in author name disambiguation in real-world scenarios.  \textbf{\# R2A} for instance, in some publications, two co-authors have the same exact names. This makes the disambiguation more difficult as these authors share not only their names but also co-authors and papers.

\begin{table}[]
    \caption{Statistical details of the top 5 sub-collections of authors sharing the same atomic name variates, where \textbf{\# ANV} is the corresponding atomic name variate, \textbf{\# UTA} is the number of unique target authors, \textbf{\# RCD} is the number of bibliographic records, \textbf{\# UCA} is the number of unique co-author full names, \textbf{\# UAN} is the number of unique target author full names, \textbf{\# R2A} is the number of records with two co-authors of the same record having the same names or the same atomic name variates and \textbf{\# R3A} is the number of records with three co-authors of the same record having the same names or the same atomic name variates. For \textbf{\# R2A} and \textbf{\# R3A}, it is not necessary that the authors have the same name / atomic name variate as the target author but most probably.}
    \label{tab:stat}
    \centering
    \begin{tabular}{|l|c|c|c|c|c|}
    \hline
         &  `Y Wang'   & `Y Zhang'   & `Y Chen'   & `Y Li'   & `Y Liu'\\
    \hline
    \textbf{\# UTA}     &  2601   & 2285   & 2260   & 2166   & 2142\\
    \hline
    \textbf{\# RCD}     &  37409  & 33639  & 26155  & 29154  & 27691\\
    \hline
    \textbf{\# UCA}     &  43199  & 39389  & 33461  & 35765  & 33754\\
    \hline
    \textbf{\# UAN}     &  2005   & 1667   & 2034   & 1734   & 1606\\
    \hline
    \textbf{\# R2A}     &  582    & 598    & 316    & 372    & 338\\
    \hline
    \textbf{\# R3A}     &  13     & 12     & 4      & 4      & 3\\
    \hline
    \end{tabular}
    \vspace{-0.3cm}
\end{table}

To ensure a credible evaluation and result reproducibility in real scenarios, we split the records in each sub-collection into a training set ($\thicksim 70\%$), validation set ($\thicksim 15\%$) and testing set ($\thicksim 15\%$) in terms of records/target author. Specifically, for each target author, we randomly split the corresponding records. If the target author did not author enough publications for the split, we prioritize the training set, then validation and finally the test set. Consequently, the number of samples is not necessarily split according to $70:15:15$ as the number of co-authors differs among publications. Moreover, it is highly likely that the records of a unique target author are completely different among the three sets. Consequently, it is difficult for the model to recognize the appropriate author only from his/her co-authors and research area. However, we believe that this is more realistic and a perfect simulation of the real scenario. 

To account for possible name variates, each input sample of full names is duplicated, where the duplicate down sample full names of all co-authors to atomic name variates. Note that this is applied to training, validation and test sets. The goal is to let the model capture all name variates for each author and his/her co-authors. In none of the sets, the variates are mixed in a single sample as we assume that this case is very less likely to occur in the real world. The experiments were conducted on a machine with the following specifications:
\begin{itemize}
    \item Processor: AMD Ryzen Threadripper 1950X 16-Core
    \item RAM: 12 GB 
    \item Graphics card: NVIDIA Titan V GV100
\end{itemize}
The algorithm was implemented in Python 3.7 using the TensorFlow library.

\subsection{Results}

The existing AND approaches use different datasets to design and evaluate their models. This lead to different assumptions and challenge disparity. Unfortunately, the codes to reproduce the results of these approaches are not available or easily accessed~\cite{hussain2017survey}. Therefore, it is not possible to fairly compare \emph{WhoIs} against baseline approaches. For future work, our code and the used datasets are publicly available~\footnote{\url{https://doi.org/10.5281/zenodo.7744775}}. 

Table~\ref{tab:result0} presents the result of \emph{WhoIs} on the sub-collections presented in Table~\ref{tab:stat}. The label \emph{All} in the table denotes that all samples were predicted twice, one with full names of the target author and its co-authors and another time with only their atomic name variates, whereas the label \emph{ANV} denotes that only samples with atomic names are predicted. The obtained results show that an important number of publications are not properly assigned to their appropriate authors. This is due to the properties of the sub-collections which were discussed above and statistically presented in Table~\ref{tab:stat}. For example, 1) two authors with the same common name authoring a single publication. 2) more than one author with the same common atomic name variate authoring a single publication, 3) number of authors with the same full name, 4) the uncertainty of the accuracy of the dataset, etc.

\begin{table}[]
    \caption{Detailed results of \emph{WhoIs} on the sub-collections corresponding to the top five of authors sharing the same atomic name variates in the DBLP repository. The results are presented in terms of Micro average precision (\textbf{MiAP}), Macro average precision (\textbf{MaAP}), Micro average recall (\textbf{MiAR}), Macro average recall (\textbf{MaAR}), Micro average F1-score (\textbf{MiAF1}) and Macro average F1-score (\textbf{MaAF1}). \textbf{ANV} denotes that only atomic name variates were used for all target authors and all their co-authors.}
    \label{tab:result0}
    \centering
    \begin{tabular}{|l|c|c|c|c|c|}

    \hline
            &  `Y Wang'   & `Y Zhang'   & `Y Chen'   & `Y Li'   & `Y Liu'\\

    \hline
    \textbf{MaAP}(ANV)&  $0.226$  & $0.212$  & $0.255$  & $0.193$  & $0.218$\\
    \textbf{MaAP}(All)&  $0.387$  & $0.351$  & $0.404$  & $0.342$  & $0.347$\\
    \hline
    \textbf{MaAR}(ANV)&  $0.299$    & $0.276$   & $0.301$   & $0.229$   & $0.267$\\
    \textbf{MaAR}(All)&  $0.433$    & $0.383$   & $0.409$   & $0.339$   & $0.361$\\
    \hline
    \textbf{MaAF1}(ANV)&  $0.239$    & $0.220$    & $0.258$     & $0.195$     & $0.223$\\
    \textbf{MaAF1}(All)&  $0.385$    & $0.342$    & $0.383$     & $0.321$     & $0.332$\\
    \hline
    \textbf{MiAF1}(ANV)&   $0.274$   & $0.278$    & $0.366$    & $0.260$    & $0.322$\\
    \textbf{MiAF1}(All)&  $0.501$    & $0.482$    & $0.561$    & $0.492$    & $0.504$\\
    \hline
    \end{tabular}
    \vspace{-0.2 cm}
\end{table}

Although the comparison is difficult and cannot be completely fair, we compare \emph{WhoIs} to other state-of-the-art approaches, whose results are reported in~\cite{zhang2017name}. These results are obtained on a collection from CiteSeerX~\footnote{\url{http://clgiles.ist.psu.edu/data/}} that contains records of authors with the name / atomic name variate `\emph{Y Chen}'. This collection consists of $848$ complete documents authored by $71$ distinct authors. We picked this name for comparison because of two reasons; 1) the number of authors sharing this name is among the top five as shown in Table~\ref{tab:stat} and 2) All methods cited in~\cite{zhang2017name} could not achieve a good result. We applied \emph{WhoIs} on this collection by randomly splitting the records into $70\%$ for training, $15\%$ for validation and $15\%$ for testing. The results are shown in Table~\ref{tab:result1}. Note that in our collection, we consider way more records and distinct authors (see Table~\ref{tab:stat}) and we use only reference attributes (i.e. co-authors, title and source). 

As the results presented in Table~\ref{tab:result1} show, \emph{WhoIs} outperforms other methods in resolving the disambiguation of the author name `\emph{Y Chen}' on the CiteSeerX dataset, which is a relatively small dataset and does not really reflect the performance of all presented approaches in real scenarios. The disparity between the results shown in Table~\ref{tab:result0} and Table~\ref{tab:result1} demonstrates that the existing benchmark datasets are manually prepared for the sake of accuracy. However, this leads to covering a very small portion of records whose authors share similar names. This disparity confirms that author name disambiguation is still an open problem in digital libraries and far from being solved.

\begin{table}[]
    \caption{Comparison between \emph{WhoIs} and other baseline methods on CiteSeerX dataset in terms of Macro F1 score as reported in~\cite{zhang2017name}. \textbf{ANV} denotes that only atomic name variates were used for all target authors and all their co-authors.}
    \label{tab:result1}
    \centering
    \begin{tabular}{|c|c|c|}
    \hline
         &  Macro ALL/ANV & Micro ALL/ANV \\
         \hline
        \emph{WhoIs} & $\mathbf{0.713}$ / $\mathbf{0.702}$ & $0.873$ / $0.861$ \\
        \hline
        NDAG~\cite{zhang2017name} & $0.367$ & N/A \\
        \hline
        GF~\cite{kuang2012symmetric} & $0.439$ & N/A\\
        \hline
        DeepWalk~\cite{perozzi2014deepwalk} & $0.118$ & N/A \\

        \hline
        LINE~\cite{tang2015line} & $0.193$ & N/A \\
        \hline
        Node2Vec~\cite{grover2016node2vec} & $0.058$ & N/A \\
        \hline
        PTE~\cite{tang2015pte} & $0.199$ & N/A \\
        \hline
        GL4~\cite{hermansson2013entity} & $0.385$ & N/A \\
        
        \hline
        Rand~\cite{zhang2017name} & $0.069$ & N/A \\
        
        \hline
        AuthorList~\cite{zhang2017name} & $0.325$ & N/A \\
        
        \hline
        AuthorList-NNMF~\cite{zhang2017name} & $0.355$ & N/A \\

        \hline
    \end{tabular}
\end{table}

The obtained results of \emph{WhoIs} illustrate the importance of relying on the research area of target authors and their co-authors to disambiguate their names. However, they trigger the need to encourage all authors to use different author identifiers such as ORCID~\cite{baglioni2021we} in their publications as the automatic approaches are not able to provide a perfect result mainly due to the complexity of the problem.

\subsection{Limitations and obstacles of \emph{WhoIs}:}
\label{subsec:limit}

\emph{WhoIs} demonstrated a satisfactory result and outperformed state-of-the-art approaches on a challenging dataset. However, the approach faces several obstacles that will be addressed in our future works. In the following, we list the limitations of the proposed approach:

\begin{itemize}[leftmargin=*]
    \item New authors cannot be properly handled by our approach, where a confidence threshold is set to decide whether the input corresponds to a new author or an existing one. To our knowledge, none of the existing supervised approaches is capable to handle this situation. 
    
    \item Commonly, authors found new collaborations which lead to new co-authorship. Our approach cannot benefit from the occurrence of new co-combinations of co-authors as they were never seen during training. \\
    \textbf{Planned solution:} We will train an independent model to embed the author's discipline using his/her known publications.  With this, we assume that authors working in the same area of research will be put close to each other even if they did not publish a paper together, the model would be able to capture the potential co-authorship between a pair of authors in terms of their area of research.
    
    \item Authors continuously extend their research expertise by co-authoring new publications in relatively different disciplines. This means that the titles and journals are not discriminative anymore. Consequently, it is hard for our approach to disambiguate authors holding common names. \\
    \textbf{Planned solution:} we plan to determine the author’s areas of research by mining domain-specific keywords from the entire paper instead of its title assuming that the author uses similar keywords/writing styles even in different research areas with gradual changes which can be captured by the model.

    \item There are a lot of models that have to be trained to disambiguate all authors in the DBLP repository. 
    
    \item Commonly, the number of samples is very small compared to the number of classes (i.e. authors sharing the same atomic name variate) which leads to overfitting the model. \\
    \textbf{Planned solution:} we plan to follow a reverse strategy of disambiguation. Instead of employing the co-authors of the target author, we will employ their co-authors aiming to find the target author among them. We aim also to learn co-author representation by employing their co-authors to help resolve the disambiguation of the target author's name.
    
    \item As mentioned earlier and stated by the maintainers of the platform~\footnote{\url{https://dblp.org/faq/How+accurate+is+the+data+in+dblp.html}}, the accuracy of the DBLP repository is not guaranteed. 
    
\end{itemize}

%% file: Sources/conclusion.tex
\section{Conclusion}
\label{conclusion}

We presented in this paper a comprehensive overview of the problem of AND. To overcome this problem, we proposed a novel framework that consists of a lot of supervised models. Each of these models is dedicated to distinguishing among authors who share the same atomic name variate (i.e. first name initial and last name) by leveraging the co-authors and the titles and sources of their known publications. The experiments on challenging and real-scenario datasets have shown promising and satisfactory results on AND. We also demonstrated the limitations and challenges that are inherent in this process.

To overcome some of these limitations and challenges, we plan for future work to exploit citation graphs so that author names can be linked to real-world entities by employing the co-authors of their co-authors. We assume that using this reverse process, the identity of the target author can be found among the co-authors of his/her co-authors. We plan also to learn the research area of co-authors in order to overcome the issue of new co-authorships.